\documentclass{jpsj3}

\def\runauthor{Yositake {\sc Takane}}
\title{%
Nonuniversal Shot Noise in Disordered Quantum Wires\\
with Channel-Number Imbalance
}

\author{%
Yositake {\sc Takane}
}

\inst{%
Department of Quantum Matter, Graduate School of Advanced Sciences of Matter,\\
Hiroshima University, Higashihiroshima, Hiroshima 739-8530, Japan
}

\recdate{ \hspace{50mm} }

\abst{%
The number of conducting channels for one propagating direction is equal to
that for the other direction in ordinary quantum wires.
However, they can be imbalanced in graphene nanoribbons with zigzag edges.
Employing the model system in which a degree of channel-number imbalance
can be controlled, we calculate the shot-noise power at zero frequency
by using the Boltzmann-Langevin approach.
The shot-noise power in an ordinary diffusive conductor
is one-third of the Poisson value.
We show that with increasing the degree of channel-number imbalance,
the universal one-third suppression breaks down and a highly nonuniversal
behavior of shot noise appears.
}

\kword{%
perfectly conducting channel, graphene nanoribbon,
Boltzmann-Langevin equation, one-third suppression
}

\begin{document}
\sloppy
\maketitle

\section{Introduction}

More than a decade ago, Barnes \textit{et al.}~\cite{barnes1,barnes2}
considered a fictitious disordered quantum wire of length $L$
having $N$ conducting channels in minority direction and
$N+m$ conducting channels in majority direction ($m = 0,1,2,\dots$),
and showed by using a scattering argument that this system has
$m$ perfectly conducting channels in the majority direction.
This indicates that if $m \ge 1$, the conductance in the minority direction
vanishes in the limit of $L \to \infty$ while that in the majority direction
converges to $(e^{2}/2\pi)m$.
That is, Anderson localization arises only in the minority direction.
Although such an intriguing behavior was predicted,
its detail has not been investigated until recently.
This is probably because a realistic example of the system
with channel-number imbalance has been lacking.

A few years ago, Wakabayashi \textit{et al.}~\cite{wakabayashi1,
wakabayashi2,wakabayashi3} have pointed out that graphene nanoribbons
with zigzag edges are a promising candidate
for the system with channel-number imbalance.
The band structure of a zigzag nanoribbon has two energy valleys, called
$K_{-}$ and $K_{+}$ valleys, well separated in momentum space,
and each valley has an excess one-way channel arising from a partially
flat band.~\cite{fujita}
Due to the presence of an excess one-way channel,
the numbers of conducting channels for two propagating directions are
imbalanced in each valley.
If impurity potentials are long-ranged and thus intervalley scattering is
absent, the two energy valleys are disconnected leading to the realization
of a channel-number imbalanced system.
Inspired by this observation, the statistical behavior of the conductance
in disordered wires with channel-number imbalance
has been studied extensively.~\cite{takane1,takane2,hirose,takane3,takane4,
takane5,kobayashi,takane6}
Recently, the present author studied the averaged conductance of this system
in the incoherent regime, where the phase coherence of conduction electrons
are completely lost, and showed that its characteristic behavior
reflecting the presence of perfectly conducting channels is unchanged
even in the absence of phase coherence.~\cite{takane7}

In this paper we study nonequilibrium shot noise in the system
with channel-number imbalance.
In an ordinary diffusive conductor with conductance $G$, the shot-noise power
$P_{\rm shot}$ at a bias voltage $V$ is given by
$P_{\rm shot} = (2/3)e|V|G$,~\cite{beenakker,nagaev}
which is one-third of the Poisson value $P_{\rm Poisson}=2e|V|G$.
This behavior, known as the one-third suppression of shot noise, has been
confirmed in several experiments.~\cite{liefrink,steinbach,schoelkopf}
The one-third suppression arises in the diffusive regime regardless of
the presence or absence of the phase coherence as long as quantum wires
are shorter than energy relaxation length.~\cite{shimizu}
Does this universality hold even in the presence of
channel-number imbalance?
To answer this question we employ the model system
in which a degree of channel-number imbalance can be controlled,
and calculate $P_{\rm shot}$ in the incoherent regime.
We show that with increasing the degree of channel-number imbalance,
the one-third suppression breaks down and a highly nonuniversal
behavior of shot noise appears.
We set $\hbar = k_{\rm B} = 1$ and ignore the spin degeneracy
throughout this paper.

\section{Model and Formulation}

We describe our model system for disordered quantum wires
with channel-number imbalance.~\cite{takane7}
As schematically shown in Fig.~1, its band structure
has two energy valleys, $K_{-}$ and $K_{+}$.
\begin{figure}[btp]
\begin{center}
\includegraphics[height=3cm]{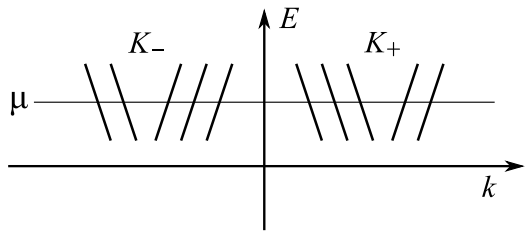}
\end{center}
\caption{The band structure of a quantum wire with two energy valleys
$K_{-}$ and $K_{+}$, in each of which the numbers of conducting channels are
imbalanced between two propagating directions.
}
\end{figure}
We assume that the numbers of conducting channels are imbalanced
in each valley between two propagating directions
although the total number of conducting channels in one direction
is equal to that in the other direction.
For concreteness we assume that the $K_{-}$ valley has $N+m$ right-moving
channels and $N$ left-moving channels, and the $K_{+}$ valley has
$N$ right-moving channels and $N+m$ left-moving channels (see Table~I).
\begin{table}[bth]
\caption{The number $N_{\rm R}$ of right-moving channels
and the number $N_{\rm L}$ of left-moving channels in each valley.
}
\begin{center}
\begin{tabular}{c|c|c}
\hline
Valley & $N_{\rm R}$ & $N_{\rm L}$ \\
\hline
$K_{-}$ & $N+m$ & $N$ \\
$K_{+}$ & $N$ & $N+m$ \\
\hline
\end{tabular}
\end{center}
\label{t1}
\end{table}
Figure~1 corresponds to the case of $N = 2$ and $m = 1$.
When $m \ge 1$, a degree of channel-number imbalance can be controlled
by the strength of intervalley scattering due to impurities.
Apparent channel-number imbalance appears if intervalley scattering is absent,
while it disappears with increasing the strength of intervalley scattering.
Let $\varepsilon_{{\rm R}nk}^{\pm}$ ($\varepsilon_{{\rm L}nk}^{\pm}$) be
the energy of an electron state with wave number $k$ in the $n$th right-moving
(left-moving) channel of the $K_{\pm}$ valley.
We introduce the corresponding distribution function $g_{Xnk}^{z}$,
where $X = {\rm R}, {\rm L}$ and $z = +,-$.
The group velocity is given by
$v_{Xnk}^{z} = \partial\varepsilon_{Xnk}^{z}/\partial k$.

We assume that our model system of length $L$ is placed along the $x$ axis,
and is connected to left and right reservoirs
at $x = -L/2$ and $L/2$, respectively.
A bias voltage $V$ is applied between the two reservoirs,
and thus the electron distribution is described by
$f_{\rm L}(\varepsilon) = f_{\rm FD}(\varepsilon - eV/2)$
in the left reservoir and
$f_{\rm R}(\varepsilon) = f_{\rm FD}(\varepsilon + eV/2)$
in the right reservoir,
where $f_{\rm FD}$ represents the Fermi-Dirac function.
We study noise power of this model on the basis of
the Boltzmann-Langevin approach.
An expression of noise power has been derived
in refs.~\citen{de_Jong1} and \citen{de_Jong2} for disordered conductors
in a continuum representation.
To adapt it to our model we must take account of its unique band structure
by introducing the concept of subbands.
We outline the derivation of noise power in a subband representation.

We introduce $\eta \equiv \{ Xnkz \}$ to characterize electron states.
For later convenience we represent the time-reversed counterpart
of $\eta$ as $\bar{\eta} \equiv \{ \bar{X}n\bar{k}\bar{z} \}$,
where $\bar{\rm R} = {\rm L}$, $\bar{\rm L} = {\rm R}$, $\bar{k} = -k$,
and $\bar{\pm} = \mp$.
We decompose $g_{\eta}(x,t)$ into the average and fluctuations as
\begin{align}
  g_{\eta}(x,t) = f_{\eta}(x)+\delta f_{\eta}(x,t) ,
\end{align}
where $f_{\eta}(x) = \langle g_{\eta}(x,t) \rangle$
with $\langle \cdots \rangle$ indicating the average over $t$.
The averaged part obeys the Boltzmann equation
in a subband representation,~\cite{akera}
\begin{align}
      \label{eq:Boltzmann_eq}
     \left(v_{\eta}\frac{\partial}{\partial x}-I_{\rm imp}\otimes
     \right) f_{\eta}(x) = 0 ,
\end{align}
under the boundary condition
$f_{\eta}(-L/2) = f_{\rm L}(\varepsilon_{\eta})$ and
$f_{\eta}(L/2) = f_{\rm R}(\varepsilon_{\eta})$.
The collision integral for impurity scattering is given by
\begin{align}
  I_{\rm imp}\otimes f_{\eta}(x)
  = \sum_{\eta'}W_{\eta,\eta'}(f_{\eta'}(x)-f_{\eta}(x))
\end{align}
with $W_{\eta,\eta'}$ being the scattering probability
between the state with $\eta' = \{X'n'k'z'\}$ and that with $\eta = \{Xnkz\}$.
The scattering probability is expressed as
\begin{align}
  W_{\eta,\eta'}
  = 2\pi M_{\eta,\eta'}
    \delta\left(\varepsilon_{\eta'}-\varepsilon_{\eta}\right)
\end{align}
with $M_{\eta,\eta'}=\langle |U_{\eta,\eta'}|^{2}\rangle_{\rm imp}$,
where $U_{\eta,\eta'}$ is the matrix element of impurity potential
and $\langle \cdots \rangle_{\rm imp}$ indicates an ensemble average.
The time-dependent part obeys
\begin{align}
      \label{eq:Boltzmann-Langevin_eq}
   \left( \frac{\partial}{\partial t}
   + v_{\eta}\frac{\partial}{\partial x}
   - I_{\rm imp}\otimes
   \right) \delta f_{\eta}(x,t)
  = \delta j_{\eta}(x,t) ,
\end{align}
where $\delta j_{\eta}$ represents the source of
stochastic fluctuations.
This has zero average, $\langle \delta j_{\eta}(x,t) \rangle = 0$,
and satisfies~\cite{kogan}
\begin{align}
      \label{eq:j_corre}
   \langle \delta j_{\eta}(x,t) \delta j_{\eta'}(x',t') \rangle
  = L \delta(x-x') \delta(t-t') J_{\eta,\eta'}(x)
\end{align}
with
\begin{align}
         \label{eq:J-def}
    J_{\eta,\eta'}(x)
 & = \delta_{\eta,\eta'}
       \sum_{\eta''} W_{\eta,\eta''}
       \left[  f_{\eta}\big( 1-f_{\eta''} \big)
             + f_{\eta''}\big( 1-f_{\eta} \big) \right]
            \nonumber \\
 & \hspace{8mm}
     - W_{\eta,\eta'}
       \left[  f_{\eta}\big( 1-f_{\eta'}\big)
             + f_{\eta'}\big( 1-f_{\eta} \big) \right] ,
\end{align}
where $f_{\eta} \equiv f_{\eta}(x)$.
The average and fluctuations of the current are expressed as
\begin{align}
 \langle I \rangle(x)  & = \frac{e}{L}\sum_{\eta}v_{\eta}f_{\eta}(x) ,
    \\
 \delta I(x,t) & = \frac{e}{L}\sum_{\eta}v_{\eta}\delta f_{\eta}(x,t) ,
\end{align}
respectively.

It is convenient to introduce the Green's function
$\mathcal{G}_{\eta,\eta'}(x,x';t)$ which satisfies
\begin{align}
      \label{eq:Green_eq}
   \left( \frac{\partial}{\partial t}
   + v_{\eta}\frac{\partial}{\partial x}
   - I_{\rm imp}\otimes
   \right) {\mathcal G}_{\eta,\eta'}(x,x';t)
   = \delta_{\eta,\eta'}\delta(x-x')\delta(t) .
\end{align}
We can express $\delta f_{\eta}$ as
\begin{align}
   \delta f_{\eta}(x,t)
 & =  \int_{-\infty}^{t}{\rm d}t' \int {\rm d}x' \sum_{\eta'}
      {\mathcal G}_{\eta,\eta'}(x,x';t-t') \delta j_{\eta'}(x',t')
         \nonumber \\
 & \hspace{-12mm}
    + \sum_{\sigma = \pm} \int_{-\infty}^{t}{\rm d}t'
      \sum_{\eta'(\sigma v_{\eta'}<0)}
      {\mathcal G}_{\eta,\eta'}(x,x';t-t') v_{\eta'}
      \delta f_{\eta'}\big(\sigma(\frac{L}{2}+\delta),t'\big) ,
\end{align}
where $\delta$ is a positive infinitesimal.
The second term is added to represent incoming fluctuations
from the left and right reservoirs.
In the reservoirs near the sample region, the incoming fluctuations obey
\begin{align}
        \label{eq:delf_corre}
   \langle \delta f_{\eta}(x,t) \delta f_{\eta'}(x',t') \rangle
  = L \delta \big(x-x'-v_{\eta}(t-t')\big) \delta_{\eta,\eta'}
    f_{X}( 1-f_{X}) ,
\end{align}
where $f_{X} = f_{\rm L}(\varepsilon_{\eta})$ for $x, x' < - L/2$
and $f_{X} = f_{\rm R}(\varepsilon_{\eta})$ for $x, x' > L/2$.
We define $T_{\eta}$ as
\begin{align}
    T_{\eta}(x)
     = \int_{0}^{\infty} {\rm d}t
       \sum_{\eta'}v_{\eta'}{\mathcal G}_{\bar{\eta'},\bar{\eta}}(x_{0},x;t)
       + \theta(x_{0}-x) ,
\end{align}
where $\theta(x)$ is the Heaviside step function
and we set $x_{0} = -L/2$.
This represents the transmission probability that an electron at $x$
with $\bar{\eta}$ arrives at $x_{0}=-L/2$
and then goes into the left reservoir.
Alternatively, noting the presence of time-reversal symmetry,
we can interpret it as the transmission probability that an electron injected
into the quantum wire from the left reservoir arrives at $x$ with $\eta$.
We can show that $T_{\eta}(x)$ satisfies eq.~(\ref{eq:Boltzmann_eq}).
The boundary condition is given by
\begin{align}
      \label{eq:BC1}
  & T_{{\rm R}nk}^{z}(-L/2) = 1 ,
        \\
      \label{eq:BC2}
  & T_{{\rm L}nk}^{z}(L/2) = 0 .
\end{align}
In terms of $T_{\eta}(x)$,
the averaged part of the distribution is expressed as
\begin{align}
         \label{eq:f-T_expression}
  f_{\eta}(x)
  =  T_{\eta}(x)f_{\rm L}(\varepsilon_{\eta})
   + \big( 1 - T_{\eta}(x) \big) f_{\rm R}(\varepsilon_{\eta}) .
\end{align}

By using eq.~(\ref{eq:f-T_expression}), the conductance defined by
$G = \lim_{V\to 0}\langle I \rangle(x)/V$ is obtained as
\begin{align}
      \label{eq:G_expression}
  G = \frac{e^{2}}{L}\sum_{\eta}v_{\eta}T_{\eta}(x)
      \left(-\frac{\partial f_{\rm FD}}{\partial \varepsilon}
             \left( \varepsilon_{\eta} \right)
      \right) ,
\end{align}
which is independent of $x$ due to the continuity of current.
We consider the noise power at zero frequency
\begin{align}
      \label{eq:def_sn_power}
   P = 2\int_{-\infty}^{\infty} {\rm d}t
       \langle \delta I(x,t) \delta I(x,0) \rangle .
\end{align}
This is also independent of $x$, so we set $x = x_{0}$
without loss of generality.
Using eqs.~(\ref{eq:j_corre})-(\ref{eq:f-T_expression}),
we can express $P$ in terms of $T_{\eta}(x)$.
After manipulation (see ref.~\citen{de_Jong2} for details),
we obtain $P \equiv P_{\rm shot}+P_{\rm th}$ with
\begin{align}
      \label{eq:sn_p_shot}
  P_{\rm shot}
  & = \frac{2e^{2}}{L} \int {\rm d}x
      \sum_{\eta}\sum_{\eta'}W_{\eta,\eta'}
      \big(T_{\bar{\eta}}(x)-T_{\bar{\eta'}}(x)\big)^{2}
      T_{\eta}(x)\big(1-T_{\eta'}(x)\big)
      \big(f_{\rm L}-f_{\rm R}\big)^{2} ,
            \\
      \label{eq:sn_p_th}
  P_{\rm th}
  & = \frac{2e^{2}}{L}
      \sum_{\eta}v_{\eta}
      \bigg[  2T_{\eta}(-\frac{L}{2})f_{\rm L}
                                  \big(1-f_{\rm L}\big)
           - T_{\eta}(-\frac{L}{2})^{2}\Big(1-T_{\bar{\eta}}(-\frac{L}{2})
                                       \Big)
              \Big[
                 f_{\rm L}\big(1-f_{\rm L}\big)
               - f_{\rm R}\big(1-f_{\rm R}\big)
              \Big]
      \bigg] ,
\end{align}
where $f_{X} \equiv f_{X}(\varepsilon_{\eta})$.
Here $P_{\rm shot}$ and $P_{\rm th}$ represent the shot-noise power and
the thermal noise power, respectively.

\section{Calculation of Noise Power}

We hereafter ignore the $k$-dependence of $T_{Xnk}^{z}(x)$
because transport properties are determined by its value at the Fermi level.
For simplicity we assume that $v_{{\rm R}nk}^{z} = v_{\rm R}$
and $v_{{\rm L}nk}^{z} = v_{\rm L}$
with $v \equiv v_{\rm R} = -v_{\rm L} > 0$.
Furthermore, we assume that the scattering probability is determined by
only the valley indexes $z$ and $z'$,
and does not depend on details of initial and final states.
Hence $M_{Xnk,X'n'k'}^{z,z'} = M^{z,z'}$.
We define $\kappa^{z,z'} = LM^{z,z'}/v^{2}$
with $\kappa^{\pm,\pm} = \kappa$ and $\kappa^{\pm,\mp} = \kappa'$.
Note that $\kappa$ and $\kappa'$ describe intravalley scattering and
intervalley scattering, respectively.
The transmission probability $T_{Xn}^{z}(x)$ does not depend on $n$
due to the assumptions presented above.
In terms of the mean free path $l_{X}^{z}(x)$
presented in ref.~\citen{takane7}, we can express
\begin{align}
        \label{eq:Trans}
   T_{Xn}^{z}(x)
   =  \frac{1}{2}-\frac{x}{L}
    + {\rm sign}(v_{X})\frac{l_{X}^{z}(x)}{L} .
\end{align}
The mean free path is given as
$l_{+}(x) \equiv l_{\rm R}^{-}(x)=l_{\rm L}^{+}(-x)$
and $l_{-}(x) \equiv l_{\rm R}^{+}(x)=l_{\rm L}^{-}(-x)$ with
\begin{align}
       \label{eq:result_l}
 l_{\pm}(x)
 & = x + \frac{L}{2}
    + \frac{L}{\Sigma}
      \bigg[ \mp m(\kappa-\kappa')\left( c(x)-c_{0} \right)
        \nonumber \\
 & \hspace{-8mm}
             + \frac{m^{2}(\kappa^{2}-{\kappa'}^{2})}{\sqrt{\alpha}}
               \left( d(x)-d_{0} \right)
             - 2(2N+m)q\kappa'c_{0}\big(x+\frac{L}{2}\big)
      \bigg] ,
\end{align}
where $\alpha=(\kappa+\kappa')[m^{2}\kappa+(8N^{2}+8Nm+m^{2})\kappa']$,
$q=(2N+m)(\kappa+\kappa')$, and
\begin{align}
  c(x) & = \frac{q}{\sqrt{\alpha}}\cosh(\sqrt{\alpha}x)
           - \sinh(\sqrt{\alpha}x) ,
             \\
  d(x) & = - \frac{q}{\sqrt{\alpha}}\sinh(\sqrt{\alpha}x)
           + \cosh(\sqrt{\alpha}x) .
\end{align}
The remaining constants are given as
$c_{0}=c(-L/2)$, $d_{0}=d(-L/2)$, and
\begin{align}
  \Sigma = 4(2N+m)\kappa'
           \Big(1 + \frac{qL}{2}\Big)c_{0}
           + 2m^{2}\frac{\kappa^{2}-{\kappa'}^{2}}{\sqrt{\alpha}}d_{0} .
\end{align}
We can show that eq.~(\ref{eq:Trans}) satisfies eq.~(\ref{eq:Boltzmann_eq})
under the boundary condition of eqs.~(\ref{eq:BC1}) and (\ref{eq:BC2}).
The difference between $l_{+}(x)$ and $l_{-}(x)$ should be regarded as
a manifestation of channel-number imbalance.

Substituting eq.~(\ref{eq:Trans}) into eq.~(\ref{eq:G_expression}) we obtain
\begin{align}
       \label{eq:res-conductance}
  G = \frac{e^{2}}{2\pi}\frac{2\alpha c_{0}}{(\kappa+\kappa')\Sigma} .
\end{align}
As shown in ref.~\citen{takane7}, the averaged conductance $G$ rapidly
converges toward $(e^{2}/2\pi)m$
with increasing $L$ at $\kappa'/\kappa = 0$.
If $\kappa'/\kappa$ is sufficiently small but finite,
$G$ rapidly decreases to $(e^{2}/2\pi)m$ in an initial stage
and then crosses over to a slow decrease toward zero.
The slow decrease is accelerated with increasing $\kappa'/\kappa$.
From eq.~(\ref{eq:sn_p_th}) we obtain $P_{\rm th} = 4GT$
which is equivalent to the result for ordinary diffusive conductors.
We next consider the shot-noise power.
Equation~(\ref{eq:sn_p_shot}) contains the summations over
$\eta = \{ Xnkz \}$ and $\eta' = \{ X'n'k'z' \}$.
Carrying out the integrations over $k$ and $k'$, we obtain
\begin{align}
       \label{eq:shot-expression}
   P_{\rm shot}
   = \frac{e^{2}F_{V,T}}{\pi}
     \int {\rm d}x \sum_{Xnz}\sum_{X'n'z'} \kappa^{z,z'}
     \left(T_{\bar{X}n}^{\bar{z}}(x)-T_{\bar{X'}n'}^{\bar{z'}}(x) \right)^{2}
     T_{Xn}^{z}(x)\left(1-T_{X'n'}^{z'}(x) \right) ,
\end{align}
where $F_{V,T} = e|V|\coth(e|V|/2T)-2T$.
Note that $F_{V,T} = e|V|$ when $e|V| \gg T$.
We define the shot-noise factor
$p_{\rm shot} \equiv P_{\rm shot}/P_{\rm Poisson}$
with $P_{\rm Poisson} \equiv 2F_{V,T}G$.
This characterizes the suppression of shot noise
and is equal to $1/3$ in ordinary diffusive conductors.
Substituting eq.~(\ref{eq:Trans}) into eq.~(\ref{eq:shot-expression})
and using eq.~(\ref{eq:res-conductance}) we obtain
\begin{align}
   p_{\rm shot}
  = \frac{(\kappa+\kappa')\Sigma}{4\alpha c_{0}L^{4}} \int {\rm d}x
    \bigg[
           & 2\kappa (N+m)N\big(l_{+}(x)+l_{-}(-x)\big)^{2}
             \Big[ L^{2} + 4\big(l_{+}(-x)+x\big)\big(l_{-}(x)-x\big) \Big]
        \nonumber \\
           & \hspace{-15mm}
           + \kappa'(N+m)^{2} \big(l_{+}(x)+l_{+}(-x)\big)^{2}
             \Big[ L^{2} + 4\big(l_{+}(x)-x\big)\big(l_{+}(-x)+x\big) \Big]
        \nonumber \\
           & \hspace{-15mm}
           + \kappa'N^{2} \big(l_{-}(x)+l_{-}(-x)\big)^{2}
             \Big[ L^{2} + 4\big(l_{-}(x)-x\big)\big(l_{-}(-x)+x\big) \Big]
        \nonumber \\
           & \hspace{-15mm}
           + 2\kappa'(N+m)N \big(l_{+}(x)-l_{-}(x)\big)^{2}
             \Big[ L^{2} - 4\big(l_{+}(-x)+x\big)\big(l_{-}(-x)+x\big) \Big]
     \bigg] .
\end{align}

We briefly examine two limiting cases without channel-number imbalance.
One is $m = 0$ with arbitrary $\kappa'/\kappa$ and the other is
$\kappa'/\kappa = 1$ with arbitrary $m$.
The mean free path is expressed as $l_{\pm}(x) = (x+L/2)/\Lambda$
with $\Lambda = 1+N(\kappa+\kappa')L$ in the first case
and $\Lambda = 1+(2N+m)\kappa L$ in the second case.
It should be noted that $l_{+}(x) = l_{-}(x)$ indicates the absence
of channel-number imbalance.
In both the cases, we obtain
\begin{align}
  p_{\rm shot} = \frac{1}{3}\left(1-\Lambda^{-3}\right)
\end{align}
when $\Lambda \gg 1$,
indicating the universal one-third suppression of shot noise.
\begin{figure}[bht]
\begin{center}
\includegraphics[height=5cm]{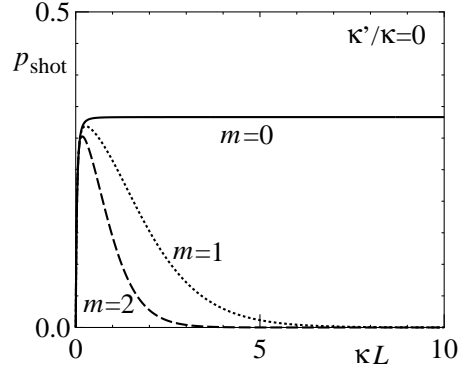}
\end{center}
\caption{$p_{\rm shot}$ as a function of $\kappa L$ in the limit of
$\kappa'/\kappa = 0$ for $N = 10$ and $m = 0$, $1$, and $2$.
}
\end{figure}
\begin{figure}[bht]
\begin{center}
\includegraphics[height=5cm]{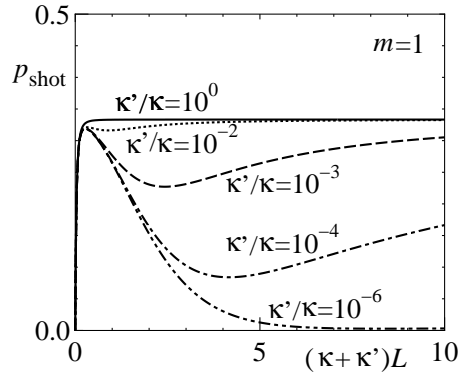}
\end{center}
\caption{$p_{\rm shot}$ as a function of $(\kappa+\kappa') L$
when $N = 10$ and $m = 1$
for $\kappa'/\kappa = 10^{0}$, $10^{-2}$, $10^{-3}$, $10^{-4}$, and $10^{-6}$.
}
\end{figure}
We present the result for $p_{\rm shot}$
in the presence of channel-channel imbalance for the case of $N = 10$.
Figure~2 displays $p_{\rm shot}$ as a function of $\kappa L$
in the no inter-valley scattering limit of
$\kappa'/\kappa = 0$ for $m = 0$, $1$, and $2$.
We observe that $p_{\rm shot}$ is strongly suppressed below the universal
value $1/3$ and the suppression becomes large with increasing $m$.
This is attributed to the fact that $G$ rapidly converges toward
$(e^{2}/2\pi)m$ with increasing $\kappa L$
while $P_{\rm shot}$ monotonically decreases.
Figure~3 displays $p_{\rm shot}$ as a function of $(\kappa+\kappa')L$
at $m = 1$ for $\kappa'/\kappa = 10^{0}$, $10^{-2}$
$10^{-3}$, $10^{-4}$, and $10^{-6}$.
The behavior of $p_{\rm shot}$ is very anomalous when $\kappa'/\kappa < 1$.
As seen at $\kappa'/\kappa = 10^{-3}$,
after an initial increase in the quasi-ballistic regime,
$p_{\rm shot}$ decreases with increasing $(\kappa+\kappa')L$
as in the case of $\kappa'/\kappa = 0$,
but then turns to increase toward the universal value of $1/3$.
This behavior reflects the fact that channel-number imbalance becomes
irrelevant in the long-$L$ limit unless $\kappa'/\kappa = 0$.
Indeed, the mean free path in this limit is approximately given by
$l_{+} = l_{-} = (x+L/2)/\Lambda$ with
\begin{align}
 \Lambda = 1+\frac{(2N+m)q\kappa'}
                  {2(2N+m)\kappa'+m^{2}\frac{\kappa^{2}-{\kappa'}^{2}}
                                             {\sqrt{\alpha}}
                                             } L
\end{align}
except for the vicinity of $x = \pm L/2$
and hence the influence of channel-number imbalance disappears.
This is consistent with the observation given below
eq.~(\ref{eq:res-conductance}).
At $\kappa'/\kappa = 0$, the difference between $l_{+}$ and $l_{-}$
never vanish and therefore channel-number imbalance
remains to be essential even in the long-$L$ limit.

\section{Summary}

We have calculated the shot-noise factor $p_{\rm shot}$,
which characterizes the suppression of shot noise compared
with the Poisson value, in disordered quantum wires
with channel-number imbalance in the incoherent regime.
We have employed the simple model in which
a degree of channel-number imbalance can be controlled
by the relative strength $\kappa'/\kappa$ of intervalley scattering.
Channel-number imbalance completely disappears at $\kappa'/\kappa = 1$,
while its influence becomes significant with decreasing $\kappa'/\kappa$.
We have shown for $m \ge 1$ that $p_{\rm shot}$ decreases
with increasing wire length $L$ after an initial increase
in the quasi-ballistic regime,
but then turns to increase toward the universal value of $1/3$
unless $\kappa'/\kappa = 0$.
The suppression of $p_{\rm shot}$ becomes large
with decreasing $\kappa'/\kappa$.
This anomalous behavior should be contrasted to
the universal one-third suppression of shot noise
widely observed in ordinary diffusive conductors.

\section*{Acknowledgment}

This work was supported in part by a Grant-in-Aid for Scientific
Research (C) (No. 21540389)
from the Japan Society for the Promotion of Science.

\end{document}